\documentclass[aps, prx, twocolumn, superscriptaddress, floatfix]{revtex4-2}
\usepackage{graphicx}
\usepackage{tikz}
\usetikzlibrary{matrix}
\usetikzlibrary{quantikz}
\usepackage{dcolumn}
\usepackage{times}
\usepackage{physics}
\usepackage{mathtools}
\usepackage{wasysym}
\usepackage{amsmath}
\usepackage{amssymb}
\usepackage{amsfonts}
\usepackage{hhline}
\usepackage{multirow}
\usepackage{hhline}
\usepackage{xcolor}
\usepackage[final]{listings}
\usepackage{hyperref}
\usepackage{graphicx}
\usepackage{soul}
\usepackage{placeins}
\usepackage{caption}

\hypersetup{colorlinks,citecolor=blue,linkcolor=blue,urlcolor=blue}
\definecolor{darkgreen}{rgb}{0,0.5,0}
\definecolor{darkred}{rgb}{0.5,0,0}
\definecolor{darkblue}{rgb}{0,0,0.5}
\definecolor{MediumGray}{gray}{0.60}

\lstdefinestyle{pseudo}{
	language=Python,
	xleftmargin=5.0ex,
	moredelim = [s][\textit]{[}{]},
	basicstyle={\small\ttfamily},
	captionpos=b,
	columns=flexible,
	numbers=left,
	numberstyle=\small,
	numbersep=8pt,
	stepnumber=1,
	numberstyle=\tiny\color{gray},
	escapechar=\%,
	breaklines=true,
	frame=single,
	framexleftmargin=15pt,
	tabsize=2,
	postbreak=\mbox{\textcolor{red}{$\hookrightarrow$}\space},
	captionpos=b,
	showspaces=false,
	showtabs=false,
	keywords=[3]{QuantumState, List, Tuple},
	commentstyle=\itshape\color{MediumGray},
}

\definecolor{electricpurple}{rgb}{0.75, 0.0, 1.0}

\begin{document}

\title{High-fidelity and Fault-tolerant Teleportation of a Logical Qubit using Transversal Gates and Lattice Surgery on a Trapped-ion Quantum Computer}

\author{C. Ryan-Anderson}
\email{ciaran.ryan-anderson@quantinuum.com}
\author{N. C. Brown}
\email{natalie.brown@quantinuum.com}
\author{C. H. Baldwin}
\author{J. M. Dreiling}
\author{C. Foltz}
\author{J. P. Gaebler}
\author{T. M. Gatterman}
\author{N. Hewitt}
\author{C. Holliman}
\author{C. V. Horst}
\author{J. Johansen}
\author{D. Lucchetti}
\author{T.~Mengle}
\author{M.~Matheny}
\author{Y. Matsuoka}
\author{K. Mayer}
\author{M. Mills}
\author{S. A. Moses}
\author{B. Neyenhuis}
\author{J. Pino}
\author{P. Siegfried}
\author{R. P. Stutz}
\author{J. Walker}
\author{D. Hayes}
\affiliation{Quantinuum, 303 South Technology Ct., Broomfield, CO 80021, USA}

\date{\today}

\begin{abstract}
Quantum state teleportation is commonly used in designs for large-scale fault-tolerant quantum computers. Using Quantinuum's H2 trapped-ion quantum processor, we implement the first demonstration of a fault-tolerant state teleportation circuit for a quantum error correction code - in particular, the planar topological $[[7,1,3]]$ color code, or Steane code. The circuits use up to $30$ trapped ions at the physical layer qubits and employ real-time quantum error correction - decoding mid-circuit measurement of syndromes and implementing corrections during the protocol. We conduct experiments on several variations of logical teleportation circuits using both transversal gates and lattice surgery protocols. Among the many measurements we report on, we measure the logical process fidelity of the transversal teleportation circuit to be $0.975\pm0.002$ and the logical process fidelity of the lattice surgery teleportation circuit to be $0.851\pm0.009$. Additionally, we run a teleportation circuit that is equivalent to Knill-style quantum error correction and measure the process fidelity to be $0.989\pm0.002$.
\end{abstract}
\maketitle
\thispagestyle{empty} 

The quantum teleportation protocol was first discovered by Bennett et al.~\cite{Bennett1993} and was experimentally demonstrated shortly thereafter~\cite{Bouwmeester1997,Boschi1998,Riebe2004,Barrett2004,Olmschenk2009}. By pre-distributing an entangled pair of qubits to two registers, a third qubit can be teleported between the registers via transferring two bits of classical information. This protocol was immediately recognized as an enabler for quantum computing or quantum networks, where qubits need to interact in non-local geometries but moving them is challenging or slow. By distributing entangled pairs throughout a quantum computer, qubits can, in principle, be transmitted at the same speed as classical information. As it is strongly believed that fault tolerance will be needed for large quantum computations, it is ultimately at the logical level~\cite{erhard2020entangling,Bluvstein2023} where teleportation becomes truly enabling.

In this work, we demonstrate the maturity of the H-series trapped-ion quantum processors by performing the first fault-tolerant version of the state teleportation circuit using a quantum error correction (QEC) code (i.e., a quantum code with distance of three or greater). We demonstrate four logical variants of the teleportation protocol using up to 30 trapped-ion qubits and real-time QEC and contrast the results with the teleportation of physical qubits. Two of the logical variants use lattice surgery to implement logical gates, which, to the best of our knowledge, are the first demonstrations of lattice surgery performed on a QEC code. 
These experiments used Quantinuum's H2 trapped-ion quantum processor~\cite{Moses2023}, which currently operates with 32 physical qubits. We give more hardware details throughout the text and in Appendix~\ref{appendix_Hardware}.

\section{Experimental Designs}
In the teleportation protocol, the task is to send a state $\ket{\psi}$ from one qubit register to another. We refer to the qubit initially encoding $\ket{\psi}$ as qubit one. To begin, an entangled Bell-pair, $(\ket{00}+\ket{11})/\sqrt{2}$, is created as a quantum resource state to facilitate the protocol, and we refer to these qubits as qubits two and three (see Fig.~\ref{fig:Teleportation}). In the next step, a joint measurement of qubits one and two is made by first entangling them and then performing individual measurements on them, effectively measuring $X \otimes X$ and $Z \otimes Z$. This measurement ``disassembles'' $\ket{\psi}$
and yields two bits of classical information. The joint measurement (or Bell measurement) reveals no local information about $\ket{\psi}$, preventing a wave-function collapse, and only reveals global information about the bit and phase parities between the measured qubits. The two bits of classical information are then used to choose appropriate single qubit operations to reassemble $\ket{\psi}$ in the unmeasured qubit three. So the state of qubit one is teleported into qubit three via the transfer of two classical bits.

This section describes the three distinct groups of teleportation experiments we performed: (1) a physical-level protocol, (2) a logical-level protocol using transversal gates, and (3) a logical-level protocol using lattice surgery gates.

\subsection{Physical-level Experiment}
We establish a performance baseline by first characterizing the teleportation protocol implemented at the physical level. To do so, we constructed circuits consisting of four independent teleportation circuits running in parallel. Four copies were run to average the performance across the four gating and measurement zones in the H2 trap~\cite{Moses2023}; therefore, the physical teleportation circuits used 12 qubits.

The teleportation circuit was benchmarked by sending an informationally complete set of states through the circuit and measuring the probability of finding the correct state at the output. The six eigenstates of the single-qubit Pauli operators were used as inputs: $\ket{\psi} \in \{\ket{0},\ket{1},\ket{+},\ket{-},\ket{+i},\ket{-i}\}$. 
The average state fidelity of the teleportation over this set is $F_a = \tfrac{1}{6}\sum_{\psi} p_{\psi}$, where $p_{\psi}$ is the probability of finding the correct state as the output. This quantity determines the process fidelity using $F_p = ((d+1) F_a - 1)/d$~\cite{Nielsen2002}, where, the dimension, $d=2$ for teleportation of a single qubit. Note, this method of fidelity estimation does not distinguish different sources of error (e.g., SPAM versus gate and memory errors). In this work, we are focused on the different teleportation protocols studied and their overall fidelity. We leave it to future work to analyze component-level performance on the physical and logical level~\cite{mayer2024benchmarking}.

\begin{figure}
\includegraphics[trim=0 0 0 0, clip, width=\columnwidth]{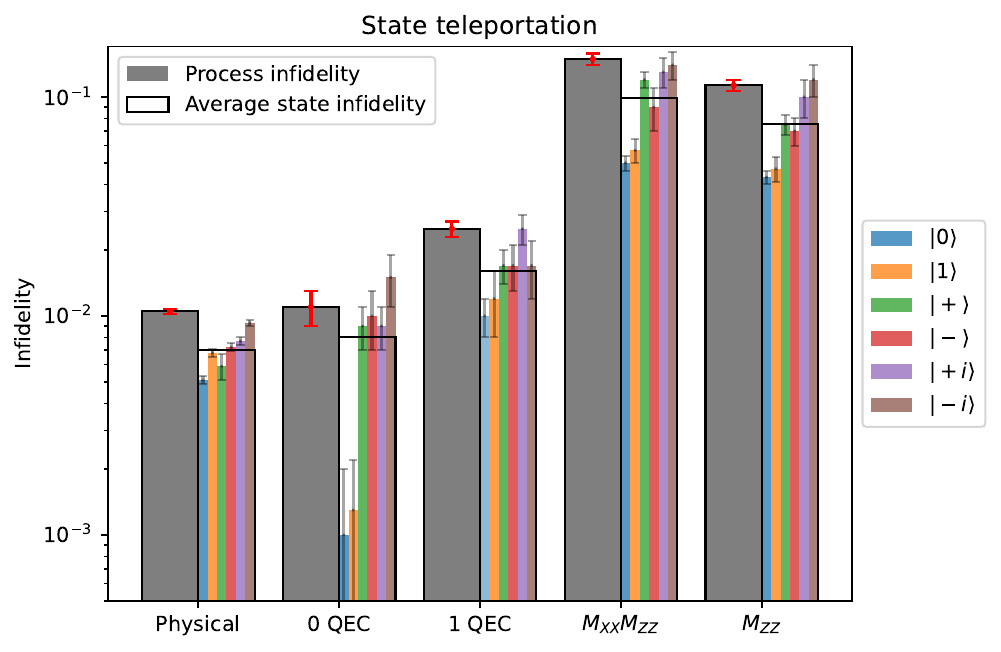}
\caption{The measured infidelities of teleportation protocol. All six eigenstates of the three Pauli operators were used as inputs to the gadget, with each state's infidelity as denoted by the individual colors. The informationally complete set of states is sufficient to estimate the average state infidelity, $1-F_a$ (clear box around individual infidelities) and process infidelity $1-F_p$ (grey box).}
\label{fig:process_fid}
\end{figure}

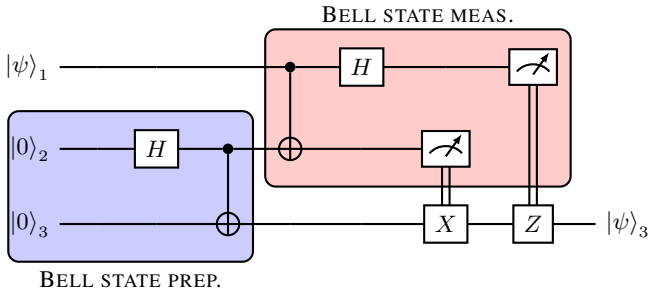
\begin{figure}
\begin{quantikz}
\ghost{}&\lstick{$\ket{\psi}_1$}& \qw   &  \qw   & \qw     &\ctrl{1}\gategroup[2,steps=4,style={rounded
corners,fill=red!20, inner
xsep=2pt},background,label style={label
position=above,anchor=north,yshift=0.2cm}]{{\sc
Bell state meas.}}&\gate{H}& \qw  & \meter{}  \vcw{2} &      \\
\ghost{}\gategroup[2,steps=5,style={rounded
corners,fill=blue!20, inner
xsep=2pt},background,label style={label
position=below,anchor=north,yshift=-0.2cm}]{{\sc
Bell state prep.}}&\lstick{$\ket{0}_2$}   & \qw   &\gate{H}&\ctrl{1} &\targ{} & \qw    &\meter{}   \vcw{1}         & &\\
\ghost{}&\lstick{$\ket{0}_3$}   & \qw   &   \qw  &\targ{}  & \qw    &  \qw   &\gate{X}                &\gate{Z} &  \qw   \rstick{$\ket{\psi}_3$} 
\end{quantikz}
\caption{The general circuit for the teleportation circuit. The highlighted sub-circuit labeled ``Bell state prep" creates the entangled Bell-pair $\ket{00}+\ket{11}$.}
 \label{fig:Teleportation}
\end{figure}

We measure the process fidelity of the physical-level experiment to be $0.9895^{+3}_{-3}$, with details in Fig.~\ref{fig:process_fid} and Tables~\ref{Table:FidelitiesCombined},\ref{Table:FidelitiesPhys}. We note that the state fidelities of the physical-level experiment are lower for the $-1$ eigenstates compared to the $+1$ states. This could be due to a known bias of the device at the physical level in measuring $\ket{0}$ with slightly higher fidelity than $\ket{1}$~\cite{qtuumspec}, but a careful investigation is beyond the scope of this work.

\renewcommand{\arraystretch}{1.5}%
\begin{table}[tp]
\resizebox{0.95\columnwidth}{!}{
\begin{tabular}{|c|c|c|c|c|c|}
\hline
\textbf{$\ket{\psi}$} & Physical           & 0 QEC             & 1 QEC             & $M_{XX}M_{ZZ}$        & $M_{ZZ}$ \\ \hline \hline
$\ket{0}$             & $0.9949^{+2}_{-2}$ & $0.999^{+1}_{-1}$ & $0.990^{+2}_{-2}$ & $0.950^{+4}_{-4}$     & $0.957^{+3}_{-3}$  \\ \hline
$\ket{1}$             & $0.9932^{+3}_{-3}$ & $0.9987^{+9}_{-9}$ & $0.988^{+4}_{-4}$ & $0.943^{+7}_{-7}$     & $0.953^{+6}_{-6}$  \\ \hline
$\ket{+}$             & $0.9941^{+8}_{-8}$ & $0.991^{+2}_{-2}$  & $0.983^{+3}_{-3}$ & $0.88^{+1}_{-1}$      & $0.925^{+8}_{-8}$  \\ \hline
$\ket{-}$             & $0.9928^{+3}_{-3}$ & $0.990^{+3}_{-3}$  & $0.983^{+4}_{-4}$ & $0.91^{+2}_{-2}$      & $0.93^{+1}_{-1}$  \\ \hline
$\ket{+i}$            & $0.9923^{+3}_{-3}$ & $0.991^{+2}_{-2}$  & $0.975^{+4}_{-4}$ & $0.87^{+2}_{-2}$     & $0.90^{+2}_{-2}$  \\ \hline
$\ket{-i}$            & $0.9907^{+3}_{-3}$ & $0.985^{+4}_{-4}$  & $0.983^{+5}_{-5}$ & $0.86^{+2}_{-2}$     & $0.88^{+2}_{-2}$  \\ \hline 
\hline
$F_a$                 & $0.9930^{+2}_{-2}$ & $0.992^{+1}_{-1}$ & $0.984^{+1}_{-1}$ & $0.901^{+6}_{-6}$     & $0.925^{+5}_{-5}$  \\ \hline
$F_p$                 & $0.9895^{+3}_{-3}$ & $0.989^{+2}_{-2}$ & $0.975^{+2}_{-2}$  & $0.851^{+9}_{-9}$     & $0.887^{+7}_{-7}$  \\ \hline
$F_{a,\mathrm{QED}}$  & NA      & $0.9997^{+2}_{-4}$  & $0.9999^{+1}_{-3}$  & $0.974^{+3}_{-3}$  & $0.995^{+1}_{-1}$ \\ \hline
$F_{p,\mathrm{QED}}$  & NA      & $0.9996^{+3}_{-5}$  & $0.9998^{+2}_{-5}$  & $0.962^{+5}_{-5}$  & $0.992^{+1}_{-1}$ \\ \hline
\end{tabular}
}
\caption{The measured fidelities of teleportation protocol. All six eigenstates of the three Pauli operators were used as inputs to the gadget, with each state's fidelity denoted as $F_s$. The informationally complete set of states is sufficient to estimate the average state fidelity $F_a$ and process fidelity $F_p$. The results obtained by analyzing the data as a QED code are labeled with a QED subscript.}
\label{Table:FidelitiesCombined}
\end{table}

\subsection{Logical-level Experiments: Transversal Circuits}
\label{subsec_trans_gates}

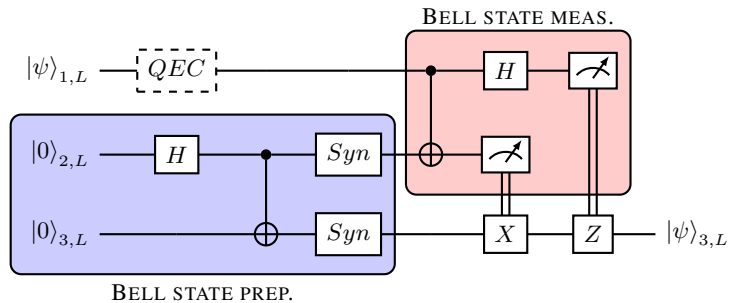
\begin{figure*}
\begin{quantikz}
\ghost{}&&\lstick{$\ket{\psi}_{1,L}$}   &  \gate[style={dashed}]{QEC}   &   \qw &\qw&\ctrl{1}\gategroup[2,steps=3,style={solid,rounded
corners,fill=red!20, inner
xsep=2pt},background,label style={label
position=above,anchor=north,yshift=0.2cm}]{{\sc
Bell state meas.}}&\gate{H}          &\meter{}  \vcw{2}        &      \\
 \ghost{}\gategroup[2,steps=6,style={solid,rounded
corners,fill=blue!20, inner
xsep=2pt},background,label style={label
position=below,anchor=north,yshift=-0.2cm}]{{\sc
Bell state prep.}}&&\lstick{$\ket{0}_{2,L}$}  &\gate{H}&\ctrl{1} &\gate{Syn}&\targ{}     &\meter{}   \vcw{1}         & &\\
\ghost{}&&\lstick{$\ket{0}_{3,L}$}     &   \qw  &\targ{}  &\gate{Syn}& \qw     &\gate{X}                &\gate{Z} & \qw  \rstick{$\ket{\psi}_{3,L}$} 
\end{quantikz}
\caption{The logical-level circuit for the teleportation circuit. The gadget labeled $\textit{``QEC''}$ is a full QEC round with up to two rounds of adaptive syndrome extraction ~\cite{RyanAnderson2021,reichardt2020fault}, and the box is dashed to highlight that it is omitted in some experiments. The sub-circuit labeled Bell state prep. shows where the entangled Bell-pair is created, which we consider a resource state. The gadgets labeled \textit{``Syn''} are each a round of flagged syndrome extraction used to post-select Bell state preparations without non-trivial syndrome measurements.}
 \label{fig:LogicalTeleportation}
\end{figure*}

This section describes two equivalent \textit{logical} teleportation protocols composed entirely of transversal gates.
In benchmarking transversal logical operations, a natural question arises -- whether or not to include QEC gadgets\footnote{Here, we define QEC gadgets as the fault-tolerant measurement of syndromes (which perform an identity operation on the logical information), but have sufficient information to infer and correct faults.}. 
It is not entirely uncommon to hear claims in the emerging field of experimental QEC that transversal circuits do not require QEC gadgets to achieve fault tolerance, but in the context of large-scale, real-world devices, this claim is oversimplified and deserves further nuance. 

The claim stems from technical definitions of fault-tolerance~\cite{Aliferis2007} and a requirement that circuit components not allow a single fault to spread into multiple faults. Transversal gates trivially satisfy this property. Since two composed transversal operations can generate another transversal operation, one might conclude there is no limit to how large a transversal circuit can be, maintaining fault tolerance without using QEC gadgets. 
However, simple probabilistic noise accumulation in a series of transversal operations will eventually overwhelm any code and introduce logical errors, setting a natural limit to the number of transversal operations that can be done without QEC gadgets.
The situation is compounded when the logical circuit connectivity becomes more complex, as this can lead to faults cascading so that what was originally a correctable fault becomes uncorrectable.

When a quantum computer's transversal operations have error rates above or near the threshold, we suggest it is prudent for benchmarking experiments to include QEC gadgets. Our logical CNOTs operate near the pseudo-threshold (our physical CNOT error is $\mathcal{O}(10^{-3})$). We include QEC gadgets in some of these circuits and look at circuits without QEC gadgets to gain information about the logical error budget. 

The teleportation circuit utilizing transversal CNOTs is illustrated in  Fig.~\ref{fig:LogicalTeleportation}.
The circuit is encoded with the $[[7,1,3]]$ QEC code, or Steane code ~\cite{steane1996error}, which has been used in many demonstrations of logically encoded circuits ~\cite{nigg2014quantum,hilder2021faulttolerant,Postler2021,Bluvstein2023}, including our previous work ~\cite{RyanAnderson2021,RyanAnderson2022}. Refs.~\cite{RyanAnderson2021,RyanAnderson2022} detail all the techniques and subroutines relevant to this section, and here we provide a brief overview of them.

The logical-level teleportation circuit is analogous to the physical-level circuit.
What is different is the underlying physical implementation necessary for encoding and performing quantum error correction. 
First, logical qubits two and three are prepared in $\ket{0}_L$ using a fault-tolerant encoding circuit, which includes a single non-destructive logical $\overline{Z}$ measurement per logical qubit to verify proper preparation of the state (we use the $L$ subscript for logical states and overlines for logical operators) ~\cite{goto2016minimizing,RyanAnderson2021, RyanAnderson2022}. Transversal Hadamard and CNOT gates are then applied to qubits two and three to create the Bell-pair ~\cite{RyanAnderson2022}.
Then, qubits two and three undergo a simple round of syndrome extraction using flagged circuitry~\cite{RyanAnderson2021}, measuring each of the six syndromes once.
As illustrated in Fig.~\ref{fig:LogicalTeleportation}, the syndrome extraction gadgets implemented on qubits two and three occur before either of those qubits interact with the qubit initially storing $\ket{\psi}_L$; therefore, the syndrome extraction gadgets are considered part of the resource Bell-state preparation.

We take advantage of the resource state construction by noting that, in large-scale fault-tolerant architectures, it is scalable to use post-selection to create resource states such as magic states or Bell pairs. Such architectures assume large devices will have capacity to prepare resource states ahead of time and in parallel, allowing well-prepared resource states to be available for the computation when they are needed~\cite{Nielsen00,knill2004faulttolerant_a,knill2004faulttolerant_b,knill2005quantum,Aliferis2007}. In this work, well-prepared Bell states are post-selected using both the verification measurement in the encoding circuit \textit{and} the syndrome measurements after the entangling operation. The scheme does not eliminate all errors from the Bell pair; for example, a weight-three fault resulting in a logical operator being applied would go undetected. 

Note, this logical Bell resource-state preparation method was first developed for this work for use in the logical teleportation protocols and was then analyzed as an individual component in Ref.~\cite{dasilva2024demonstration}. A description of the performance of this logical Bell state preparation component can be found in the reference.

After the Bell state is prepared, logical qubit one is initialized to $\ket{0}_L$ using the encoding circuit~\cite{goto2016minimizing,RyanAnderson2021}. In contrast to the preparation of the Bell-state resource, the encoding circuit for logical qubit one uses up to three rounds of a repeat-until-success protocol~\cite{RyanAnderson2021}, at which point the circuit proceeds regardless of the outcome of the verification step. That is, we do not post-select on the preparation of logical qubit one. After preparing $\ket{0}_L$, a transversal single-qubit Clifford operation, natively admitted by the Steane code, can prepare one of the six states, $\{\ket{0}_L,\ket{1}_L,\ket{+}_L,\ket{-}_L,\ket{+i}_L,\ket{-i}_L\}$.

We implemented two variants of the logical circuit depicted in Fig.~\ref{fig:LogicalTeleportation}. These variants differ only by the presence/absence of a full QEC gadget on logical qubit one. 
Given the importance of QEC gadgets in large-scale fault-tolerant computation, we consider the benchmarking with the QEC gadget included as important. However, examining the circuit without the QEC gadget offers insight into the logical error contributions of the teleportation protocol. Note, we call the protocol that includes one QEC gadget on logical qubit one as ``1QEC'' and the protocol without the QEC gadget as ``0QEC.'' Interestingly, the entire 0QEC experiment can be viewed as the implementing a QEC gadget as originally proposed by Knill~\cite{knill2004scalable,knill2005quantum}.

In the 1QEC experiment, the QEC gadget implemented is the full conditional adaptive syndrome extraction circuitry described in Ref.~\cite{RyanAnderson2021}. The extracted syndromes are processed in real-time using a look-up table decoder.
Active error correction is then applied, which involves physically applying Pauli corrections (instead of Pauli frame tracking).

Next, we perform the Bell measurement on qubits one and two.
All the data qubits in logical qubits one and two are measured destructively during the Bell measurement. These measurements are used to reconstruct syndrome information, decode, and correct the logical outcomes. 
The corrected outcomes are then used to update the Pauli frame of qubit three, equivalent to conditionally applying the classically controlled  Pauli gates on qubit three.

To measure the probability of finding the correct state in logical qubit three (i.e., determine the state fidelity), we apply a single-qubit transversal gate to measure in a particular basis, destructively measure all data qubits, reconstruct syndromes, decode, and apply corrections to the final logical outcome (see~\cite{RyanAnderson2021} for details).

For clarity, we emphasize that real-time decoding occurs four times in this circuit. During the QEC gadget and the three destructive logical measurements, the decoder measures syndromes and determines corrections.

We estimate the process fidelities of these circuits to be $0.975^{+2}_{-2}$ when a QEC gadget is included on logical qubit one and $0.989^{+1}_{-1})$ when the QEC gadget is omitted. This result suggests the QEC gadget is one of the noisier components of the circuit, consistent with previous measurements~\cite{RyanAnderson2021,RyanAnderson2022}. There was also a difference in the total number of physical qubits used in each experiment, with the included QEC gadget version using 30 physical qubits (10 for each code block) and the version without a QEC gadget requiring 28 (10 for each code block in the Bell state, and for the other code block, 7 data qubits and an additional verification ancilla for the encoding circuit).  

We further re-evaluated the data considering the framework of quantum error detection (QED). In this analysis, in addition to post-selecting on Bell-state preparation, we also post-select based on non-trivial syndrome information obtained from the QEC gadget and the outcomes of destructive logical measurements. This approach offers data for work studying QED distance-two codes to compare against the performance of QEC codes when viewed through the lens of QED. Moreover, this analysis sheds additional light on the underlying noise model by indicating the rate at which noise induces pure logical errors, which are undetectable and uncorrectable, during computation. Establishing an upper bound on performance, if error rates derived from re-assessing the data as a QEC code are notably high, this signals a potential need to revisit the experimental setup. When processing the data as a QED code, we find process fidelities of $0.9996_{-5}^{+3}$ for 0QEC and $0.9998_{-5}^{+2}$ for 1QEC. Further details are given in Fig.~\ref{fig:process_fid} and Appendix~\ref{appendix_data}.

\subsection{Logical Experiment: Lattice surgery}
\label{latt_surg}

\begin{figure*}
\centering
\includegraphics[trim=105 610 95 66, clip, scale=1.05]{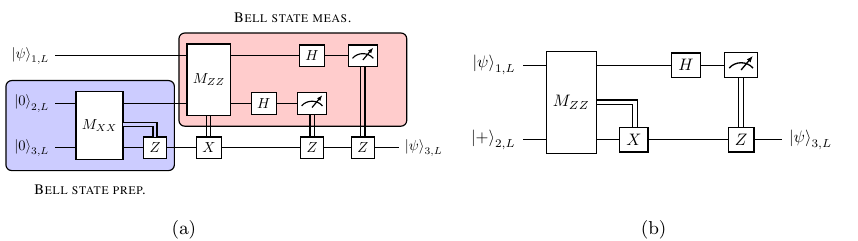}
\caption{Two different implementations of logical teleportation circuits using a lattice surgery gate set. The first circuit (a) is analogous to the transversal implementation in Fig.~\ref{fig:LogicalTeleportation}. The second circuit (b) is logically equivalent (see Fig.~\ref{fig:lattice_surg_proof} in Appendix~\ref{appendix:circuit_ids}) but does not use a Bell resource state and utilizes less logical qubits.}
\label{fig:latt_surg_circuit}
\end{figure*}

Transversal operations are not the only methods for performing logical gates. In this section, we demonstrate the flexibility of the quantum charge-coupled device (QCCD) architecture~\cite{Wineland98} by looking at equivalent logical teleportation circuits built from a color-code lattice-surgery gate set~\cite{landahl2014quantum}.
Lattice surgery is a strategy for implementing logical gates that require only 2D nearest-neighbor interactions, making it especially useful for architectures whose qubit locations are fixed in a 2D space ~\cite{Horsman2012, landahl2014quantum}. 
The gate set of lattice surgery is made up of the joint measurements of logical Pauli operators. 
These joint operators are measured when merging code blocks together into a joint code block by measuring stabilizers between and within the original code blocks~\cite{Horsman2012, landahl2014quantum}. 
Unlike their transversal counterparts, lattice surgery protocols inherently require some form of syndrome measurements, with fault-tolerant implementation requiring a full QEC gadget. 
Since QEC gadgets are done \textit{with} the logical gate implementation, direct comparisons with transversal gate sets are complicated.
Lattice surgery circuits have previously been demonstrated in ion-traps ~\cite{erhard2021entangling} using $d=2$ quantum error detecting surface codes, and we build on this work by demonstrating these techniques on a $d=3$ quantum error-correcting color code.

An analogous logical teleportation circuit using lattice surgery techniques can be seen in Fig.~\ref{fig:latt_surg_circuit}. Some methods used here are identical to those used in the transversal circuits, such as the initialization protocols of all three logical qubits, transversal Hadamard gates, decoding, and destructive measurements.

The Bell-state preparation uses lattice-surgery gate sets.
After qubits two and three have been initialized, a joint $\overline{X_2X_3}$ logical operator is measured (Fig.~\ref{fig:joint_pauliXX} in appendix Sec.~\ref{appendix:circuit_ids}).
This measurement scheme includes an additional physical flag qubit to catch higher-weight hook errors ~\cite{dennis2002topological, yoder2017surface, RyanAnderson2021}. 
After the measurement of the joint logical operator, we perform one round of syndrome extraction using flagged checks (also used in the ``Syn" gate in Fig.~\ref{fig:LogicalTeleportation}). 
We again treat the Bell state as a resource state and post-select on the same criteria as the transversal implementation (encoding verification failing and any non-trivial syndromes) and any non-trivial measurement from the flag qubit in the $\overline{X_2X_3}$ joint logical measurement.
For this experiment, we only post-select on the Bell-state preparation. 
We decode the syndrome information in real-time, using the same decoder as the other experiments, and keep track of corrections on qubit three via a Pauli frame. 

The circuit proceeds by measuring a second joint logical operator $\overline{Z_1Z_2}$, again using a flag circuit (Fig.~\ref{fig:joint_pauliZZ} in appendix Sec.~\ref{appendix:circuit_ids}). 
We then perform a full QEC gadget with both the flagged and conditional unflagged syndrome measurements. 
If the logical measurement flag or any syndromes in the QEC gadget are non-trivial, we repeat the measurement of $\overline{Z_1Z_2}$, followed by the ``Syn'' gadget syndrome measurements (Fig.~\ref{fig:LogicalTeleportation}). 
We treat this second measurement round as ideal and continue with the circuit. 
If the first round of measurements of $\overline{Z_1Z_2}$ and the full QEC gadget yield trivial flags and syndromes, the second round of measurements is skipped. 
The rest of the circuit involves transversal Hadamard gates, destructive measurements, decoding, and conditional corrections tracked via a Pauli frame, all of which are handled in the same way as in the transversal teleportation circuits. 

The lattice-surgery circuit, which we call $M_{XX}M_{ZZ}$, is measured to have a process fidelity of $0.851^{+9}_{-9}$, and when further post-selected as a QED code, we find a process fidelity of $0.962^{+5}_{-5}$, (details in Appendix~\ref{appendix_data}). These QED process fidelities lag behind those of the transversal two-qubit gate QEC process fidelities, suggesting that enhancements in decoding strategies alone will not suffice for the lattice surgery protocols to achieve parity with the performance of the transversal protocols.
A total of 30 physical qubits were used in this experiment: 10 for each individual code block. 
The depth of the lattice surgery circuit is greater than that of its transversal counterpart, which likely contributes to the decreased fidelity.

Lattice surgery offers more than an alternative way to implement logical gates, it can also reduce overheads in some cases ~\cite{litinski2019game}. 
In this spirit, we also investigated an equivalent logical teleportation circuit using only two logical qubits instead of three. 
This version does not use a Bell state resource but can be shown to be algorithmically equivalent to teleportation protocol.
A ``proof", showing the equivalency of this simpler circuit (Fig.~\ref{fig:latt_surg_circuit}(b)) to the typical teleportation circuit (Fig.~\ref{fig:LogicalTeleportation}), can be found in the Appendix~\ref{appendix:circuit_ids} (see Fig.~\ref{fig:lattice_surg_proof}). This form of lattice surgery teleportation can be seen as merging a logical qubit with an ancillary code block and then shrinking the newly created merged code block back to what was the auxiliary code block, transferring the logical state to the logical ancilla. The protocol itself can be viewed as a lattice surgery version of 1-bit teleportation~\cite{Zhou2000},
This technique can be used for various things, such as code-switching and gate teleportation ~\cite{poulsen2017fault}. It was previously experimentally implemented for a d=2 QED code in Ref.~\cite{erhard2021entangling}; here, we implement it for a QEC code utilizing real-time decoding and error correction.

The simpler teleportation circuit can be seen in Fig.~\ref{fig:latt_surg_circuit} (b). 
The beginning of the circuit is similar to the other experiments, using encoding circuits with verification ancilla implemented in a repeat-until-success protocol to encode $\ket{0}_L$. 
The repeat-until-success circuit is performed up to three times for both logical qubits to achieve high-fidelity input states.
From there, the joint Pauli $\overline{Z_1Z_2}$ logical operator is measured, followed by a full QEC gadget measuring both the flagged and conditional unflagged syndrome extraction protocol.
The rest of the circuit, including decoding, correction tracking via Pauli frame, transversal single qubit gates, and destructive measure outs, are performed like all the other experiments. 

We measure this lattice surgery circuit, which we call $M_{ZZ}$, to have a process fidelity of $0.887^{+7}_{-7}$ and when further post-selected as a QED code, $0.992^{+1}_{-1}$. 
Only 20 qubits were needed in this implementation, 10 for each individual code block.
Further details are given in Fig.~\ref{fig:process_fid} and Appendix~\ref{appendix_data}.

\section{Discussion}
\label{sec_results}
\label{sec_disc}

The main results of all the experiments are in Fig.~\ref{fig:process_fid} and Table~\ref{Table:FidelitiesCombined} (see Appendix~\ref{appendix_data} for additional details). The data in Fig.~\ref{fig:process_fid} reflects the asymmetric noise environment of the H-series quantum computers. As Ref.~\cite{RyanAnderson2021} notes, the circuit error budget is dominated by two physical layer noise sources - two-qubit gate errors and memory errors. While the gate errors are not completely without structure, we find that a depolarizing error model usually captures their impact on typical circuits. However, memory error is strongly dominated by stochastic and coherent rotations about $Z$. The CSS structure of the $[[7,1,3]]$ code promotes the physical $Z$ errors to the logical level since the logical $\overline{Z}$ operator is a product of physical $Z$ operators (Fig. 1 in Ref.~\cite{RyanAnderson2021}). The eigenstates of logical $\overline{X}$ and $\overline{Y}$ are sensitive to unintended logical $\overline{Z}$ operations and, therefore, yield lower fidelity. Note that the physical layer fidelities do not show the same asymmetry because gate errors dominate the physical circuits more than the logical circuits.

The process fidelity of the physical teleportation experiment is estimated at $0.9895^{+3}_{-3}$, setting a high bar for achieving the break-even point with logical circuits. The logical transversal experiment without the QEC gadget on logical qubit one exhibits a process fidelity of $0.989^{+2}_{-2}$, suggesting a break-even milestone. However, it would be more conclusive if logical fidelity surpassed physical fidelity with reasonable statistical separation. Considering our previous work, which is near break-even with transversal two-qubit gates for the Steane code~\cite{RyanAnderson2022}, achieving break-even in this context could be plausible. A more stringent and significant break-even milestone for logical teleportation involves comparing the physical circuits to the corresponding logical experiments with mid-circuit QEC gadgets since they are key to preventing logical errors and fault propagation. Doing so, we find the process fidelity of $0.975^{+2}_{-2}$ to be slightly lower than the corresponding physical process fidelity of $0.9895^{+3}_{-3}$. We are optimistic that even without further hardware improvements, this process fidelity can be improved by further optimizations at the logical level, as discussed below.

The experiments 0QEC, 1QEC, and MXXMZZ can be viewed as implementing the Knill-style QEC gadget~\cite{knill2004scalable,knill2005quantum}. The 0QEC version closely mirrors Knill's original proposal. For Knill-style QEC, $X$ and $Z$ syndromes are assessed during teleportation by destructively measuring logical qubits one and two. These measurements determine logical outcomes and gather syndromes data, which are subsequently decoded. These corrections generated are then not only used to teleport the state but also to fix accumulated faults.

Knill-style QEC is particularly interesting as it supports a wide class of codes capable of logical teleportation, avoids generating hook error when transversal two-qubit gates are used, reduces non-Pauli noise such as leakage, and offers single-shot decoding. This last feature trades reduced time overheads for the requirement of two logical ancillas.

The promising fidelity of $0.989^{+2}_{-2}$ for the Knill gadget observed in 0QEC suggests its performance. We can calculate the infidelity (one minus the process fidelity) as $1.1\mathrm{e}{-2}$ and roughly compare the error rate of QEC gadget used in 1QEC by subtracting the process fidelities of 0QEC from 1QEC. Doing so, we estimate a logical error rate of $1.4\mathrm{e}{-2}$, potentially increased by memory errors induced on the other logical qubits during its operation. This indicates the two gadgets may have similar performance; however, future studies should conduct a detailed component-level analysis to properly compare the performance of the different QEC gadget implementations.

Despite these encouraging preliminary estimates, we believe the current results do not definitively confirm whether we have met or exceeded a break-even point for the QEC gadgets. We suggest two informative break-even milestones for assessing QEC gadgets' effectiveness: 1) ensuring the gadget's error rate is at or below that of the physical gate most limiting the performance of general physical algorithms (typically the two-qubit gate) and 2) comparing the error rate of the physical two-qubit gate to that of the fault-tolerant logical two-qubit gate when QEC gadgets are employed. However, to hit break-even milestones at the algorithmic level, logical gates must achieve error rates and order of magnitude lower or more due to the increased count of logical gates. This stems from the discrete sets of protected gates that QEC codes utilize~\cite{Nielsen00,Eastin09,dawson2005solovaykitaev}.

One notable contribution of this work is the introduction of lattice surgery techniques in the QCCD architecture. Our initial experiments in lattice surgery showed lower process fidelities than the transversal circuits, but we stress that the lattice surgery circuits likely have room for more optimization. One should not draw strong conclusions about the prospects for lattice surgery versus transversal methods in QCCD devices as the H2 design was not optimized for either method, and less time was spent considering how to optimize the lattice surgery protocols. We hope to continue to improve our experimental implementation of these and other logical procedures.

Optimizing the complex structure of the logical circuits was made easier by programming on the logical level utilizing a domain-specific language known as SLR (see Appendix~\ref{appendix_Hardware}). Other straightforward optimizations may include different but logically equivalent QEC protocols, reducing leakage errors~\cite{hayes2020eliminating,suchara2015leakage,brown2020critical}, Clifford deformation~\cite{Tuckett_2018,Bonilla_Ataides_2021,Vasmer_2022}, dynamical decoupling, Pauli twirling and other circuit-level techniques to suppress coherent noise~\cite{debroy2018stabilizer, zhang2021hidden, Parrado_Rodr_guez_2021}, larger distance codes~\cite{iverson2019coherence}, and improved decoding. The decoders in this work only consider the syndromes and determine corrections for individual subcircuits, such as the QEC gadget and logical measurements. Further improvements in decoding may be possible by incorporating knowledge of the bias noise present in the system as well as by decoding over the collection of all syndromes measured in the circuit~\cite{Bacon2017,Gottesman2022,Delfosse2023,Bluvstein2023}.

The results in this work represent the state-of-the-art in experimental QEC; however, more work is needed to demonstrate the error suppression promised by QEC. Fortunately, we believe we are far from any fundamental limitations. The largest noise source in the two-qubit gate is spontaneous emission, which can be reduced using different laser wavelengths and power~\cite{Ozeri07}. Memory error, especially any coherent component, is also of special concern to the performance of deep logical programs. We note that our results were generated using the software stack available to any user of Quantinuum’s H-series computers and ran alongside customer jobs. The software stack compiler is optimized for general purposes, not specific error correction circuits. 
Thus, the results in this work represent typical performance currently seen for the H2-1 machine.
Custom compilation techniques could reduce transport times and the associated memory error. Also, QCCD ion-traps using 2D geometries~\cite{Delaney2024} could significantly increase the clock speed. 

Fault tolerance and error suppression come with an overhead price, and understanding the amount of tolerance needed, at a cost of circuit depth, width, and time, compared to the gain in error suppression is an important consideration as quantum devices scale up.
To this end, we demonstrate the first logical state teleportation and the first instance of lattice surgery for a QEC code, demonstrating the ability of the QCCD technology to explore different QEC paradigms. In that same spirit, another trapped-ion quantum processor recently demonstrated code-switching~\cite{Pogorelov2024}, which could be an enabling technique for non-Clifford operations and should be suitable for QCCD architectures. 
We expect many more studies in the emerging field of experimental QEC to be done to optimize resource requirements, including resource state generation and use, gate set comparisons, and investigating \textit{practical} fault-tolerance beyond idealized mathematical definitions. Through investigations like these, quantum hardware and tailored QEC protocols may be co-designed to accelerate the progress toward large-scale quantum computing.

\section{Acknowledgements}

We acknowledge the large team at Quantinuum for all their contributions and the fabrication facility at Honeywell for producing world-class ion traps. We’d also like to thank Ben
Criger for helpful discussions, Jonhas Colina for helping develop the H2-1 compiler, and the good people at IARPA for inspiring discussions. Lastly, we would like to thank Tony Uttley for his leadership, support, and friendship.

\bibliography{inversion}

\appendix

\section{The H2 quantum processor}
\label{appendix_Hardware}
H2 is based on the QCCD architecture~\cite{Wineland98} and, therefore, uses ion-transport operations to achieve full-connectivity. This is a key enabler for some of the QEC protocols in this work, such as transversal entangling operations and logical joint Pauli measurements~\cite{RyanAnderson2021,RyanAnderson2022}. H2 uses $^{171}$Yb$+$ ions~\cite{Olmschenk2007} for physical qubits and $^{138}$Ba$+$ ions for sympathetic cooling, using a total of 64 ions in the experiments described in this work. The physical qubits are gated via stimulated Raman transitions implemented with off-resonant lasers directed at four different gating regions. At the time of these experiments, randomized benchmarking~\cite{Magesan2011} experiments averaged over all four gate zones showed fidelities of $\sim3\times10^{-5}$ and $1.4\times10^{-3}$ for single-qubit and two-qubit gates, respectively. SPAM errors are measured to be $\sim2\times10^{-3}$, and the crosstalk errors from mid-circuit SPAM operations are $<10^{-4}$. Memory errors are more difficult to characterize in the QCCD architecture as different circuits require different transport sequences. Previously reported benchmarking experiments~\cite{Moses2023} showed that memory errors are smaller than the gate error for certain types of circuits, but the impact of memory error on the circuits used in this work has not yet been fully characterized.

The experiments were designed at the logical level using an internally developed domain specific language for QEC dubbed Simple Logical Representation (SLR)~\cite{pecos}. SLR is used to construct libraries of QEC protocols, build logical programs from theses libraries, and resolve these logical protocols into physical circuits. This greatly simplifying the manual labor involved as well as debugging and optimization efforts. After designing the experiments in this work using SLR and generating the physical-level circuits, they were then submitted to the H-series compiler stack to produce machine-level instructions for ion-transport operations and laser pulses~\cite{Pino2020}.

\section{Additional Experimental Data}
\label{appendix_data}

\FloatBarrier

This appendix includes additional experimental details for the different teleportation experiment variants.

In this manuscript, we have focused more on the QEC data instead of the QED data since QEC is more pertinent to the development of large-scale fault-tolerant computation. Aggressive forms of QED applied to QEC programs might be useful as a potential bridge between the NISQ and fault-tolerant eras, but if discard rates are too high the utility of such techniques will be limited. For example, when viewed as a QED code, the transversal logical circuit that used the QEC gadget achieves a process fidelity of $0.9998^{+2}_{-5}$, but approximately half of the shots had to be discarded (Table~\ref{Table:Fidelities1QECQED}). Regardless, it can be useful to consider the space-time trade off of using QEC protocols in a QED manner compared to other QED protocols that aim to bridge between the NISQ era and large-scale fault-tolerant computation. In particular circumstances, post-selection and QED can be used in a scalable manner for fault-tolerant protocols such as the generation of resource states~\cite{Nielsen00,knill2004faulttolerant_a,knill2004faulttolerant_b,knill2005quantum,Aliferis2007} or when rare but uncorrectable errors are detected such as when using even distance codes~\cite{prabhu2021distancefour}.

The shots for each experimental settings were split amongst several jobs which were randomly mixed. Most error bars indicate one standard deviation and were determined using jackknife resampling~\cite{efron1982jackknife} of the fidelities for a set of related jobs, helping to account for variation in the noise (drift) over time. When an experimental setting produced no errors for any shots (seen only for the QED experiments), jackknife sampling is not used, and the ``rule of three''~\cite{Eypasch} is employed to provide a one standard deviation lower bound. The rule of three estimates 95\% confidence intervals, so for consistency with the other errors bars we alter the rule to the analogous bound for one standard deviation. For our parenthetical error bar notation, a single number indicates a symmetric error bar around the mean, and two numbers indicates an asymmetric error bar around the mean.

Table~\ref{Table:FidelitiesPhys} contains the data for the physical qubit teleportation experiments. Table~\ref{Table:Fidelities0QECQEC} contains the data for the logical transversal teleportation experiments operating as a QEC code, and no QEC gadget is applied to the logical input state (logical qubit 1). Table~\ref{Table:Fidelities1QECQEC} contains the data for the logical transversal teleportation experiments operating as a QEC code, and one QEC gadget is applied to the logical input state (logical qubit 1). Table~\ref{Table:Fidelities0QECQED} contains the data for the logical transversal teleportation experiments operating as a QED code, and no QEC gadget is applied to the logical input state (logical qubit 1). Table~\ref{Table:Fidelities1QECQED} contains the data for the logical transversal teleportation experiments as a QED code, and one QEC gadget is applied to the logical input state (logical qubit 1). Table~\ref{Table:FidelitiesMZZ} contains the data for the logical teleportation experiments that use a lattice surgery $M_{ZZ}$ operation to teleport the logical qubit.

 \begin{table}[]
\resizebox{0.45\columnwidth}{!}{
\begin{tabular}{|c|c|c|}
\hline
\textbf{$\ket{\psi}$}   & \textbf{$F_s$} & Shots\\ \hline \hline
$\ket{0}$      & $0.9949^{+2}_{-2}$       &      $104,000$       \\ \hline
$\ket{1}$      & $0.9932^{+3}_{-3}$       &      $100,000$      \\ \hline
$\ket{+}$      & $0.9941^{+8}_{-8}$       &      $104,000$     \\ \hline
$\ket{-}$      & $0.9928^{+3}_{-3}$       &      $104,000$      \\ \hline
$\ket{+i}$     & $0.9923^{+3}_{-3}$       &      $92,000$       \\ \hline
$\ket{-i}$     & $0.9907^{+3}_{-3}$       &      $92,000$      \\ \hline \hline
$F_a$          & $0.9930^{+2}_{-2}$       &      $99,333.3$      \\ \hline
$F_p$          & $0.9895^{+3}_{-3}$       &       NA    \\ \hline
\end{tabular}
}
\caption{The measured fidelities of the physical layer teleportation gadget. All six eigenstates of the three Pauli operators were used as inputs to the gadget with each state's fidelity denoted as $F_s$. The informationally complete set of states is sufficient to estimate the average state fidelity $F_a$, and process fidelity $F_p$.}
\label{Table:FidelitiesPhys}
\end{table}

\begin{table}[]
\resizebox{0.6\columnwidth}{!}{
\begin{tabular}{|c|c|c|c|}
\hline
\textbf{$\ket{\psi}_L$}   & \textbf{$F_s$} & Accepted & Discard\\ 
   &  & shots & fraction\\ 

\hline \hline
$\ket{0}_L$      & $0.999^{+1}_{-1}$      & $1,550$ &      $0.354$       \\ \hline
$\ket{1}_L$      & $0.9987^{+9}_{-9}$     & $1,565$ &       $0.348$      \\ \hline
$\ket{+}_L$      & $0.991^{+2}_{-2}$       & $1,594$ &      $0.336$     \\ \hline
$\ket{-}_L$      & $0.990^{+3}_{-3}$       & $1,529$ &      $0.363$      \\ \hline
$\ket{+i}_L$     & $0.991^{+2}_{-2}$       & $1,538$  &     $0.359$       \\ \hline
$\ket{-i}_L$     & $0.985^{+4}_{-4}$       & $1,574$ &       $0.344$      \\ \hline 
\hline
$F_a$            & $0.992^{+1}_{-1}$      & $1558.3$ &    $0.351$      \\ \hline
$F_p$            & $0.989^{+2}_{-2}$       & NA &      NA    \\ \hline

\end{tabular}
}
\caption{The measured fidelity for the logical transversal teleportation circuit operating as a QEC code with no QEC gadget acting on the input qubit (logical qubit 1). We list the individual state fidelities $F_s$ for our set of input states $\ket{\psi}_L$, the average state fidelity $F_a$, and process fidelity $F_p$. We also list the individual discard rates associated with each input state. The discarded events are heralded by the measurement of non-trivial syndromes in the Bell state preparation sequence as described in the main text.}
\label{Table:Fidelities0QECQEC}
\end{table}

\begin{table}[]
\resizebox{0.55\columnwidth}{!}{
\begin{tabular}{|c|c|c|c|}
\hline
\textbf{$\ket{\psi}_L$}   & \textbf{$F_s$} & Accepted & Discard\\ 
   &  & shots & fraction\\ 

\hline \hline
$\ket{0}_L$      & $0.990^{+2}_{-2}$       & $1,956$ &      $0.348$       \\ \hline
$\ket{1}_L$      & $0.988^{+4}_{-4}$       & $1,997$ &      $0.346$      \\ \hline
$\ket{+}_L$      & $0.983^{+3}_{-3}$       & $1,949$ &      $0.350$     \\ \hline
$\ket{-}_L$      & $0.983^{+4}_{-4}$       & $2,015$ &      $0.328$      \\ \hline
$\ket{+i}_L$     & $0.975^{+4}_{-4}$       & $1,896$ &      $0.368$       \\ \hline
$\ket{-i}_L$     & $0.983^{+5}_{-5}$       & $1,951$ &      $0.350$      \\ \hline 
\hline
$F_a$            & $0.984^{+1}_{-1}$       & $1,960.7$ &       $0.346$      \\ \hline
$F_p$            & $0.975^{+2}_{-2}$       & NA  &      NA    \\ \hline

\end{tabular}
}
\caption{The measured fidelities of the logical transversal teleportation circuit operating as a QEC code with a QEC gadget acting on the input qubit (logical qubit 1). We list the individual state fidelities $F_s$ for our set of input states $\ket{\psi}_L$, the average state fidelity $F_a$, and process fidelity $F_p$. We also list the discard rates associated with each input state. The discarded events are heralded by the measurement of non-trivial syndromes in the Bell state preparation sequence as described in the main text.}
\label{Table:Fidelities1QECQEC}
\end{table}

\begin{table}[]
\resizebox{0.6\columnwidth}{!}{
\begin{tabular}{|c|c|c|c|}
\hline
\textbf{$\ket{\psi}_L$}   & \textbf{$F_{s, \mathrm{QED}}$} & Accepted & Discard\\ 
   &  & shots & fraction\\ 

\hline \hline
$\ket{0}_L$             & $1.0000_{-9}^{+0}$    & $1,273$    & $0.470$   \\ \hline
$\ket{1}_L$             & $1.0000_{-9}^{+0}$     & $1,288$     & $0.463$   \\ \hline
$\ket{+}_L$             & $1.0000_{-9}^{+0}$    & $1,319$    & $0.450$   \\ \hline
$\ket{-}_L$             & $0.9992_{-8}^{+8}$     & $1,273$     & $0.470$   \\ \hline
$\ket{+i}_L$            & $1.0000_{-9}^{+0}$     & $1,250$     & $0.479$   \\ \hline
$\ket{-i}_L$            & $0.9992_{-8}^{+8}$     & $1,292$     & $0.462$   \\ \hline 
\hline
$F_{a,\mathrm{QED}}$    & $0.9997_{-4}^{+2}$    & $1,282.5$  & $0.466$      \\ \hline
$F_{p,\mathrm{QED}}$    & $0.9996_{-5}^{+3}$    & NA        & NA    \\ \hline

\end{tabular}
}
\caption{The measured fidelities of the logical transversal teleportation circuit when treated as a QEC with no QEC gadget acting on the input qubit (logical qubit 1). To treat as a QED code, all non-trivial syndromes or state-preparation validation measurements are post-selection on. We list the individual state fidelities $F_s$ for our set of input states $\ket{\psi}_L$, the average state fidelity $F_a$, and process fidelity $F_p$. We also list the discard rates associated with each input state.}
\label{Table:Fidelities0QECQED}
\end{table}

\begin{table}[]
\resizebox{0.6\columnwidth}{!}{
\begin{tabular}{|c|c|c|c|}
\hline
\textbf{$\ket{\psi}_L$}   & \textbf{$F_{s, \mathrm{QED}}$} & Accepted & Discard\\ 
   &  & shots & fraction\\ 

\hline \hline
$\ket{0}_L$             & $1.0000_{-8}^{+0}$    & $1,449$    & $0.517$   \\ \hline
$\ket{1}_L$             & $1.0000_{-7}^{+0}$    & $1,549$    & $0.484$   \\ \hline
$\ket{+}_L$             & $1.0000_{-8}^{+0}$    & $1,462$    & $0.513$   \\ \hline
$\ket{-}_L$             & $1.0000_{-7}^{+0}$    & $1,535$    & $0.488$   \\ \hline
$\ket{+i}_L$            & $0.9993_{-7}^{+7}$    & $1,408$    & $0.531$   \\ \hline
$\ket{-i}_L$            & $1.0000_{-8}^{+0}$    & $1,499$    & $0.500$   \\ \hline 
\hline
$F_{a,\mathrm{QED}}$    & $0.9999_{-3}^{+1}$    & $1,483.7$  & $0.505$      \\ \hline
$F_{p,\mathrm{QED}}$    & $0.9998_{-5}^{+2}$    & NA        & NA    \\ \hline

\end{tabular}
}
\caption{Logical transversal teleportation circuit fidelities analyzed as a QED code for experiments using one QEC gadget on the input qubit (logical qubit 1). To treat as a QED code, all non-trivial syndromes or state-preparation validation measurements are post-selection on. We list the individual state fidelities $F_s$ for our set of input states $\ket{\psi}_L$, the average state fidelity $F_a$, and process fidelity $F_p$. We also list the discard rates associated with each input state.}
\label{Table:Fidelities1QECQED}
\end{table}

\begin{table}[]
\resizebox{0.6\columnwidth}{!}{
\begin{tabular}{|c|c|c|c|}
\hline
\textbf{$\ket{\psi}_L$}   & \textbf{$F_{s}$} & Accepted & Discard\\ 
   &  & shots & fraction\\ 

\hline \hline
$\ket{0}_L$             & $0.950^{+4}_{-4}$        & $2,330$    & $0.403$   \\ \hline
$\ket{1}_L$             & $0.943^{+7}_{-7}$        & $1,018$     & $0.321$   \\ \hline
$\ket{+}_L$             & $0.88^{+1}_{-1}$         & $2,321$    & $0.405$   \\ \hline
$\ket{-}_L$             & $0.91^{+2}_{-2}$         & $797$     & $0.336$   \\ \hline
$\ket{+i}_L$            & $0.87^{+2}_{-2}$        & $800$     & $0.333$   \\ \hline
$\ket{-i}_L$            & $0.86^{+2}_{-2}$        & $797$     & $0.336$   \\ \hline 
\hline
$F_{a}$    & $0.901^{+6}_{-6}$        & $1,343.8$  & $0.356$      \\ \hline
$F_{p}$    & $0.851^{+9}_{-9}$        & NA        & NA    \\ \hline

\end{tabular}
}
\caption{The measured fidelities of the $M_{XX}M_{ZZ}$ lattice surgery logical teleportation circuit when treated as a QEC code. We list the individual state fidelities $F_s$ for our set of input states $\ket{\psi}_L$, the average state fidelity $F_a$, and process fidelity $F_p$. We also list the discard rates associated with each input state. The discarded events are heralded by the measurement of non-trivial syndromes in the Bell state preparation sequence as described in the main text.}
\label{Table:FidelitiesMXXMZZQEC}
\end{table}

\begin{table}[]
\resizebox{0.6\columnwidth}{!}{
\begin{tabular}{|c|c|c|c|}
\hline
\textbf{$\ket{\psi}_L$}   & \textbf{$F_{s,  \mathrm{QED}}$} & Accepted & Discard\\ 
   &  & shots & fraction\\ 

\hline \hline
$\ket{0}_L$             & $0.980^{+2}_{-2}$        & $1,123$    & $0.712$   \\ \hline
$\ket{1}_L$             & $0.979^{+5}_{-5}$        & $562$     & $0.625$   \\ \hline
$\ket{+}_L$             & $0.982^{+3}_{-3}$         & $1,250$    & $0.679$   \\ \hline
$\ket{-}_L$             & $0.994^{+3}_{-3}$         & $455$     & $0.620$   \\ \hline
$\ket{+i}_L$            & $0.95^{+2}_{-2}$        & $476$     & $0.603$   \\ \hline
$\ket{-i}_L$            & $0.96^{+1}_{-1}$        & $450$     & $0.625$   \\ \hline 
\hline
$F_{a,  \mathrm{QED}}$    & $0.974^{+3}_{-3}$        & $719.3$  & $0.644$      \\ \hline
$F_{p,  \mathrm{QED}}$    & $0.962^{+5}_{-5}$        & NA        & NA    \\ \hline

\end{tabular}
}
\caption{The measured fidelities of the $M_{XX}M_{ZZ}$ lattice surgery logical teleportation circuit when treated as a QED code. We list the individual state fidelities $F_s$ for our set of input states $\ket{\psi}_L$, the average state fidelity $F_a$, and process fidelity $F_p$. We also list the discard rates associated with each input state. The discarded events are heralded by the measurement of non-trivial syndromes in the Bell state preparation sequence as described in the main text.}
\label{Table:FidelitiesMXXMZZQED}
\end{table}

\begin{table}[]
\resizebox{0.4\columnwidth}{!}{
\begin{tabular}{|c|c|c|}
\hline
\textbf{$\ket{\psi}_L$}   & \textbf{$F_s$} & Shots\\ 
\hline \hline
$\ket{0}_L$      & $0.957^{+3}_{-3}$      & $2,400$ \\ \hline
$\ket{1}_L$      & $0.953^{+6}_{-6}$       & $2,400$ \\ \hline
$\ket{+}_L$      & $0.925^{+8}_{-8}$       & $2,400$ \\ \hline
$\ket{-}_L$      & $0.93^{+1}_{-1}$       & $2,400$ \\ \hline
$\ket{+i}_L$     & $0.90^{+2}_{-2}$       & $2,400$ \\ \hline
$\ket{-i}_L$     & $0.88^{+2}_{-2}$       & $2,400$ \\ \hline 
\hline
$F_a$            & $0.925^{+5}_{-5}$      & $2,400$ \\ \hline
$F_p$            & $0.887^{+7}_{-7}$       & NA \\ \hline

\end{tabular}
}
\caption{The measured fidelities of the logical $M_{ZZ}$ lattice surgery gadget when treated as a QEC code. We list the individual state fidelities $F_s$ for our set of input states $\ket{\psi}_L$, the average state fidelity $F_a$, and process fidelity $F_p$. We also list the discard rates associated with each input state.}
\label{Table:FidelitiesMZZ}
\end{table}

\begin{table}[]
\resizebox{0.6\columnwidth}{!}{
\begin{tabular}{|c|c|c|c|}
\hline
\textbf{$\ket{\psi}_L$}   & \textbf{$F_{s, \mathrm{QED}}$} & Accepted & Discard\\ 
   &  & shots & fraction\\ 

\hline \hline
$\ket{0}_L$             & $0.991^{+3}_{-3}$    & $1,229$    & $0.488$   \\ \hline
$\ket{1}_L$             & $0.994^{+4}_{-4}$    & $1,269$    & $0.471$   \\ \hline
$\ket{+}_L$             & $0.9985^{+9}_{-9}$   & $1,353$    & $0.436$   \\ \hline
$\ket{-}_L$             & $0.999^{+1}_{-1}$    & $1,365$    & $0.431$   \\ \hline
$\ket{+i}_L$            & $0.993^{+2}_{-2}$    & $1,349$    & $0.438$   \\ \hline
$\ket{-i}_L$            & $0.992^{+2}_{-2}$    & $1,285$    & $0.465$   \\ \hline 
\hline
$F_{a,\mathrm{QED}}$    & $0.995^{+1}_{-1}$    & $1,308.3$  & $0.455$      \\ \hline
$F_{p,\mathrm{QED}}$    & $0.992^{+1}_{-1}$    & NA        & NA    \\ \hline

\end{tabular}
}
\caption{The measured fidelities of the logical $M_{ZZ}$ lattice surgery gadget when treated as a QED code. We list the individual state fidelities $F_s$ for our set of input states $\ket{\psi}_L$, the average state fidelity $F_a$, and process fidelity $F_p$. We also list the discard rates associated with each input state.}
\label{Table:FidelitiesMZZQED}
\end{table}


\clearpage
\onecolumngrid

\section{Circuits and identities}
\label{appendix:circuit_ids}
Here, we give explicit circuits for measuring the joint Pauli operators used in the lattice surgery experiments, a visual proof of the circuit equivalency of the simpler lattice surgery experiment to the transversal logical teleportation circuit, and include useful gate identities to help the reader.

\begin{figure*}[th]
\includegraphics[trim=0 650 0 4, clip, scale=0.85]{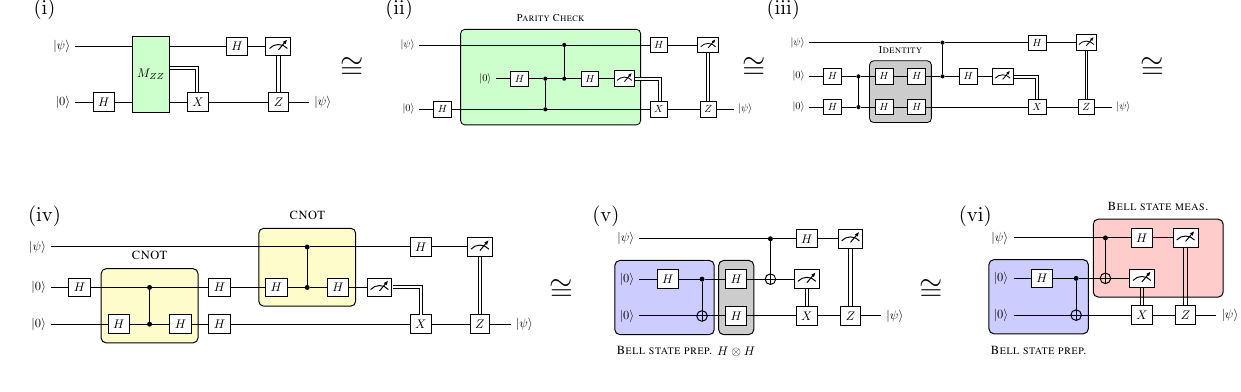}
\caption{Proof of the equivalency of the lattice surgery implementation of the teleportation circuits to that of the typical teleportation circuit.   
Starting with (i), which is exactly the circuit implemented in the experiment (see Sec.~\ref{subsec_trans_gates}), we use the relationship in Fig. ~\ref{fig:parity_check_eq} to replace the lattice surgery $M_{ZZ}$ operation with a parity check relation involving an additional ancilla qubit, to get the circuit (ii).
To get (iii), we insert two pairs Hadamards that equal the identity operation.
Using gate identities, in (iv) we highlight the $CZ$ and Hadamard gates that are equivalent to $CNOT$ gates.
Now it is clear that the lower first part of the circuit is the usual Bell pair preparation circuit which is highlighted by the blue box in (v).
The key observation here is that the $H \otimes H$ operator \textit{stabilizes} the Bell pair, and thus acts as the identity. 
Removing this set of gates, we arrive at (vi), the typical teleportation circuit illustrated in Fig. ~\ref{fig:Teleportation}.}
\label{fig:lattice_surg_proof}
\end{figure*}

\begin{figure*}[h]
\includegraphics[trim=78 705 100 50, clip, scale=0.65]{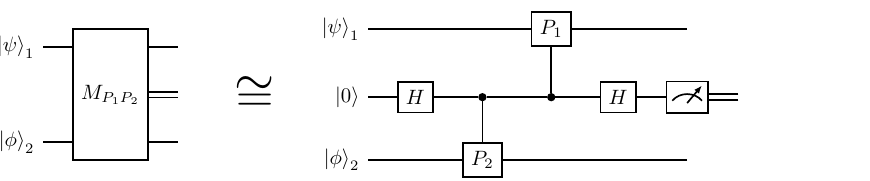}
\caption{Circuit identity for a joint Pauli measurement using an ancilla (see Ref.~\cite{Nielsen00} pg. 188 or Ref.~\cite{lidar2013quantum} pg. 72).}
\label{fig:parity_check_eq}
\end{figure*}

\begin{figure*}[th]
\includegraphics[clip, scale=1.2]{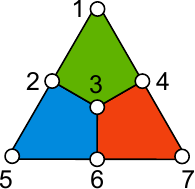}
\caption{Depiction of the Steane code. The logical representatives we use for our implementations of the logical circuits are logical operators that run along the boundary, e.g., $\overline{X} = X_5 X_6 X_7$ and $\overline{Z} = Z_5 Z_6 Z_7$. See Refs.~\cite{RyanAnderson2021,RyanAnderson2022} for more details on the Steane code and additional details on protocols we use for this code.}
\label{fig:steane_numbered}
\end{figure*}

\begin{figure*}
\begin{quantikz}
\lstick{$\ket{\psi_5}_{a,L}$} &\qw & \targ{} & \qw &\qw      &\qw     & \qw & \qw &\qw &\qw&\qw&\qw\\
\lstick{$\ket{\psi_6}_{a,L}$} &\qw & \qw     & \qw &\qw      &\targ{} & \qw & \qw &\qw &\qw&\qw&\qw\\
\lstick{$\ket{\psi_7}_{a,L}$} &\qw & \qw     & \qw &\qw      &\qw     & \qw & \targ{}&\qw &\qw&\qw&\qw\\
\lstick{$\ket{\psi_5}_{b,L}$} &\qw & \qw     & \qw & \targ{} &\qw     & \qw & \qw &\qw&\qw&\qw&\qw\\
\lstick{$\ket{\psi_6}_{b,L}$} &\qw & \qw     & \qw &\qw      &\qw     & \targ{}  &\qw &\qw&\qw&\qw&\qw\\
\lstick{$\ket{\psi_7}_{b,L}$} &\qw & \qw     & \qw &\qw      & \qw    & \qw      &\qw & \qw& \targ{}&\qw &\qw\\
\\
\lstick{$\ket{0}$} & \gate{H} & \ctrl{-7}    & \ctrl{1} &\ctrl{-4} & \ctrl{-6} & \ctrl{-3} & \ctrl{-5} & \ctrl{1} &\ctrl{-2} &\gate{H}&\meter{}\\
\lstick{$\ket{0}$} & \qw & \qw          & \targ{}  &\qw &\qw  &\qw & \qw &\targ{} &\qw  &\qw&\meter{}  \\
\end{quantikz}
\caption{The circuit used to measure the joint $\overline{X_aX_b}$ Pauli operator. Since this circuit is used in both lattice surgery implements, we label logical qubit blocks "a" and "b", to designate the different logical qubits. The subscript number indicates the index of the physical qubit within the code block. A depiction of the Steane code and corresponding numberings of the qubits can be found in Fig.~\ref{fig:steane_numbered}.}
 \label{fig:joint_pauliXX}
\end{figure*}

\begin{figure*}
\begin{quantikz}
\lstick{$\ket{\psi_5}_{a,L}$} &\qw & \ctrl{7} & \qw &\qw      &\qw     & \qw & \qw &\qw &\qw&\qw&\qw\\
\lstick{$\ket{\psi_6}_{a,L}$} &\qw & \qw     & \qw &\qw      &\ctrl{6} & \qw & \qw &\qw &\qw&\qw&\qw\\
\lstick{$\ket{\psi_7}_{a,L}$} &\qw & \qw     & \qw &\qw      &\qw     & \qw & \ctrl{5}&\qw &\qw&\qw&\qw\\
\lstick{$\ket{\psi_5}_{b,L}$} &\qw & \qw     & \qw & \ctrl{4} &\qw     & \qw & \qw &\qw&\qw&\qw&\qw\\
\lstick{$\ket{\psi_6}_{b,L}$} &\qw & \qw     & \qw &\qw      &\qw     & \ctrl{3}  &\qw &\qw&\qw&\qw&\qw\\
\lstick{$\ket{\psi_7}_{b,L}$} &\qw & \qw     & \qw &\qw      & \qw    & \qw      &\qw & \qw& \ctrl{2}&\qw &\qw\\
\\
\lstick{$\ket{0}$} & \gate{H} & \ctrl{-7}    & \ctrl{1} &\ctrl{-4} & \ctrl{-6} & \ctrl{-3} & \ctrl{-5} & \ctrl{1} &\ctrl{-2} &\gate{H}&\meter{}\\
\lstick{$\ket{0}$} & \qw & \qw          & \targ{}  &\qw &\qw  &\qw & \qw &\targ{} &\qw  &\qw&\meter{}  \\
\end{quantikz}
\caption{The circuit used to measure the joint $\overline{Z_aZ_b}$ Pauli operator. Since this circuit is used in both lattice surgery implements, we label logical qubit blocks "a" and "b", to designate the different logical qubits. The subscript number indicates the index of the physical qubit within the code block.  A depiction of the Steane code and corresponding numberings of the qubits can be found in Fig.~\ref{fig:steane_numbered}.}
 \label{fig:joint_pauliZZ}
\end{figure*}

\end{document}